\documentclass[journal]{IEEEtran}

\ifCLASSINFOpdf
  \usepackage[pdftex]{graphicx}
   \DeclareGraphicsExtensions{.pdf,.jpeg,.png}
\else
\fi

\usepackage[ruled,vlined]{algorithm2e}

\begin{document}
\title{An Efficient SDN Architecture for Smart Home Security Accelerated by FPGA}
\author{\IEEEauthorblockN{Holden Gordon, Conrad Park, Bhagyashri Tushir, Yuhong Liu, Behnam Dezfouli} \\
\IEEEauthorblockA{Internet of Things Research Lab, Computer Science and Engineering, Santa Clara University, USA\\
Emails:\{hgordon, cspark, btushir, yhliu, bdezfouli\}@scu.edu} }

\vspace{-6mm}
\maketitle
\begin{abstract}
With the rise in Internet of Things (IoT) devices, home network management and security are becoming complex. There is an urgent requirement to make smart home network management efficient.
This work proposes an SDN-based architecture to secure smart home networks through K-Nearest Neighbor (KNN) based device classifications and malicious traffic detection.
The efficiency is further enhanced by offloading the computation-intensive KNN model to Field Programmable Gate Arrays (FPGA), which offers parallel processing power of GPU platforms at lower costs and higher efficiencies, and can be used to accelerate time-sensitive tasks.
The proposed parallelization and implementation of KNN on FPGA are achieved by using the Vivado Design Suite from Xilinx and High-Level Synthesis (HLS).
When optimized with 10-fold cross-validation, the proposed solution for KNN consistently exhibits the best performances on FPGA when compared with four alternative KNN instances (i.e., 78\% faster than the parallel Bubble Sort-based implementation and 99\% faster than the other three sorting algorithms).
Moreover, with 36,225 training samples, the proposed KNN solution classifies a test query with 95\% accuracy in approximately 4 milliseconds on FPGA compared to 57 seconds on a CPU platform.
\end{abstract}

\begin{IEEEkeywords}
FPGA, smart home, IoT, security, HLS, KNN
\end{IEEEkeywords}

\IEEEpeerreviewmaketitle
\section{Introduction}

The Internet of Things (IoT) marketplace has experienced an exponential increase over the past few years. 
The current estimate projects that 43 billion IoT devices will be in circulation by 2025, a threefold increase from 2018~\cite{cisco}. Further, the investment in IoT technology is projected to grow at 13.6$\%$ per year through 2022 ~\cite{cisco}. 
One important IoT application is a smart home, whose market is expected to reach \$121 billion by 2022 \cite{bujari2018standards}. 

Existing security solutions are becoming inadequate to manage smart home IoT devices due to the following reasons. 
Firstly, IoT devices' explosive growth has led to pervasive device heterogeneity and various communication protocols, which require comprehensive network management.
Secondly, IoT devices generate massive amounts of data, resulting in heavy computational complexity for data-driven security analysis. However, the access points (APs) in most smart homes are far from adequate to handle such complex network management and analysis. Thus, an efficient and scalable security solution is required. 


To combat growing difficulties in smart home security, Software Defined Networking (SDN) technology is leveraged in this paper as an effective centralized solution. With the decoupling of the control plane and data plane, resource-limited home APs can focus mainly on simple switch functionalities on the data plane while offloading the complex control tasks. This enables SDN to provide an effective and efficient networking management system.
Given the resource-constrained nature of home gateways and routers, the proposed solution uses flow statistics as features because they are less computationally expensive than packet-by-packet statistics. Furthermore, Machine Learning (ML) models can be utilized to identify devices and malicious behaviors in a network. In particular, the K-Nearest Neighbors (KNN) algorithm is chosen in this work due to its effectiveness and simplicity. 

Despite the benefits of using ML and SDN, the capacity of SDN controllers is still limited. 
With rapid increases in the number of IoT devices in a smart home and their significant amount of traffic data, the computational overhead of running ML models is also dramatically increasing, causing challenges in real-time secure home network management.
Field Programmable Gate Arrays (FPGAs) is proposed in this paper to enhance system efficiency. They are highly flexible and can perform ML workloads with significantly reduced latency due to a fundamentally different execution paradigm than traditional CPU-based systems.
Although they can be challenging to program and utilize, FPGAs promise to enhance SDN scalability by performing auxiliary SDN tasks faster, especially those required by ML.
The proposed solution is implemented using high-level synthesis (HLS) and optimized for an FPGA workflow while comparing it to existing FPGA and parallel KNN implementations.
It is further compared to CPU-based solutions executed on a single controller to show the advantages of leveraging the network programmability of FPGA architecture to reduce the latency of ML applications on SDN.


Our contributions are listed as follows.
\textit{First}, we propose a SDN-based architecture to secure a smart home network using \textit{device classification and Distributed Denial of Service (DDoS) detection} and accelerate this decision-making process for real-time security via FPGA.
\textit{Second}, we propose a novel KNN parallelization and implementation on FPGA using HLS. 
\textit{Third}, comprehensive experimental results on both FPGA and CPU platforms validate that the proposed KNN instance outperforms other existing KNN instances in achieving the minimum latency.

The rest of this paper is organized as follows.  
We present the proposed SDN-based architecture for smart home security and data collection setup in Section \ref{ref:proposed_testing}. The overview of the KNN's implementation using HLS for device classification and DDoS detection is presented in Section \ref{ref:knn_on_fpga}. Next, the in-depth results are discussed in Section \ref{ref:results}. Finally, the related works are discussed in Section \ref{ref:related_work}, followed by a conclusion in Section \ref{ref:conclusion}.

\section{An Architecture for Smart Home Security}
\label{ref:proposed_testing}

\subsection{Proposed Architecture}
\begin{figure}[!t]
\includegraphics[width=3.5in]{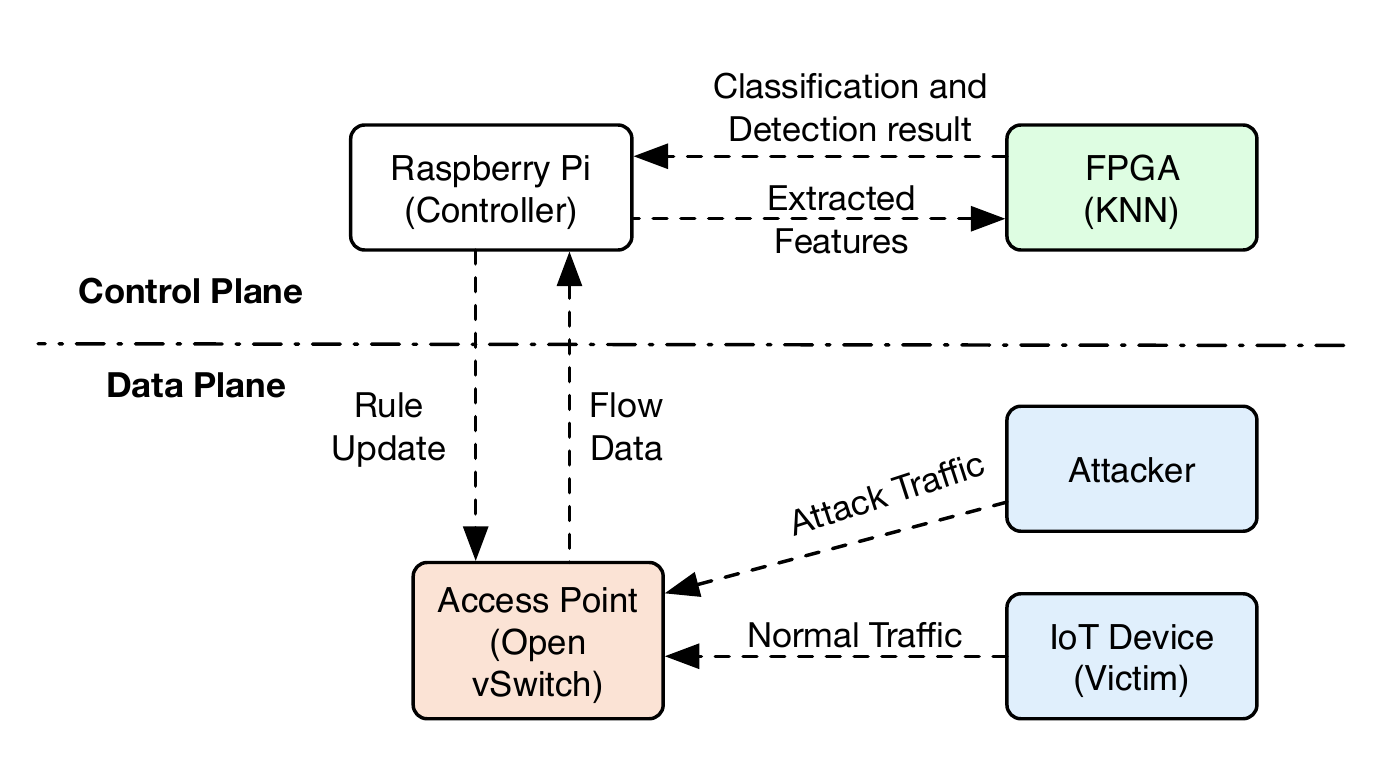}
\caption{Proposed smart home security architecture is SDN-based and accelerated by FPGA}
\label{prop_arct}
\end{figure}

We propose a SDN-based, FPGA-accelerated smart home architecture in Figure \ref{prop_arct}. The data plane comprises Open vSwitch (OVS). 
The OVS is responsible for forwarding packets as flow data to the controller received from connected devices, including IoT devices and attack devices. Further, the controller extracts features for offloading and sends them to the FPGA, where the specialized KNN implementation executes. The controller is a Raspberry Pi (RPi) running RYU controller software. The FPGA then forwards the device classification and DDoS detection class to the RPi controller to update OVS rules. Currently, the FPGA is an Artix-7 and is separated from the controller. However, this can also be realized in a single heterogeneous system that provides programmable FPGA logic and a CPU (as in the Xilinx 7000 series chips), reducing the communication lag between the FPGA and CPU and providing a complete System on Chip (SoC) that is fully integrated.

\vspace{-2mm}
\subsection{Smart Home Testing Environment}

We establish a smart home testbed to collect benign and DDoS attack data by adopting various hardware and software tools \cite{gordon2021securing}. Four popular IoT devices, including Google Home, Amazon Echo, Nest Camera, Ring Camera, and one Android phone, are adopted. 
Three Linux machines are configured to act as WiFi AP, sniffer, an attacker, respectively. 
The {\fontfamily{pcr}\selectfont hostapd} is configured on the AP. 
{\fontfamily{pcr}\selectfont hostapd} is a user-space demon that helps to create an isolated environment for data collection.
We configure {\fontfamily{pcr}\selectfont tshark} on the sniffer to collect benign and attack traffic. {\fontfamily{pcr}\selectfont tshark} is a command-line tool that reduces the graphical user interface (GUI) overhead on Linux machines. 
Further, {\fontfamily{pcr}\selectfont hping3} is installed on the attacker, which helps to change the IP source addresses, IP destination addresses, attack types, and attack duration dynamically.

The IoT devices and Android phone connect to the AP using  WiFi. 
The AP, sniffer, and attacker connect to the Ethernet switch.  We interact with IoT devices for 24 hours to collect benign traffic while the sniffer captures the traffic in pcap files. 
These interactions include various real-time activities that can occur during regular device usage.

We launch three types of DDoS attacks against IoT devices, namely ICMP, TCP-SYN, and UDP floods. 
While the sniffer records the attack traffic's pcap files, the attacker launches each type of attack for 10 minutes against the individual IoT device. This process produces a dataset of 3,066,585 packets, composed of 1,264,392 malicious packets and 1,802,193 benign packets. 

To confirm the machine learning model's robustness, we also adopt the UNSW dataset ~\cite{UNSW} consisting of 802,580 benign packets. The dataset consists of the following ten smart home devices --- Wemo Motion Sensor and Power Switch, Samsung and Netatmo Camera, TP-Link Plug, Hue Bulb, Amazon Echo, Chrome Cast, iHome, and Lifx lightbulb. 
We combine both benign datasets, i.e., our dataset (SIOTLAB) and UNSW lab. Further, the DDoS traffic is joined with the benign dataset. In our previous work ~\cite{gordon2021securing}, we identified 20 seconds as the optimal polling interval when KNN achieves accuracy greater than 95\% for device identification and DDoS detection using the following novel feature set. (1) \textit{Protocol percentage}: for each flow data, ICMP, TCP, and UDP percentage is calculated for both benign and DDoS traffic. (2) \textit{IP diversity ratio}: calculated as the number of unique IP addresses divided by the total packets sent by a device.  (3) \textit{Packet count and size}.

\section{KNN Implementation on FPGA} \label{ref:knn_on_fpga}

\subsection{KNN}
KNN is simple and easy to implement for the following reasons. 
Firstly, during the training phase, the data is stored in memory, so there is no need to build the model. Secondly, KNN has one hyperparameter to tune.
However, KNN gets significantly slower as the dataset size increases.
To overcome this challenge, we identify the minimum training samples necessary to achieve accuracy greater than 95\%.

An instance of KNN calculates distances from each of the elements in an input training dataset to a query from the testing dataset and classifies the query by returning the mode of the sorted \emph{k} nearest neighbors, where \emph{k} is the hyperparameter.
A query is classified by finding the mode or most common nearest neighbor. 
This process can be described by several tasks: calculating all of the neighbors' distances, finding the closest neighbors, and evaluating the mode of these neighbors.

The closest neighbors are determined based on a distance metric evaluated on the query against each of the training data points. We utilize the Manhattan distance metric to avoid computationally intense operations such as square roots. 
Next each data points is sorted based on the distance from the query.
This step is the most computationally-intensive task of KNN because it may require sorting the entire input dataset. 
This results in an algorithmic complexity that is linear with the size of the data multiplied by the number of features. 


FPGAs have the potential to greatly reduce the run time of KNN by running the tasks in parallel. 
Pipelining and unrolling the distance calculations for the training data points and sorting the \emph{k} nearest neighbors, it is possible to drastically shorten the classification process.


\vspace{-2mm}
\subsection{Sorting Algorithms used by KNN}
In order to thoroughly test the extent of current FPGA optimization techniques, we implement the KNN algorithm using High Level Synthesis (HLS) and evaluate the performance of several common sorting algorithms including Bubble Sort, Merge Sort, Odd-Even Sort, and Enumeration Sort. These sorting algorithms are compared against the proposed custom algorithm (K-Min Sort).

\begin{algorithm}[!tp]
\SetAlgoLined
\KwIn{The training data $train$, their labels $train\_labels$, and a test query $query$.}
\KwOut{The classification of the test query.}
$distances[k]\gets\{\infty\}$\;
$labels[k]\gets\{\}$\;
 
\ForEach{$e\in train$ and $l\in train\_labels$}{%
    $tmp\_dist\gets$ distance$(e, query)$\;
    $tmp\_max\gets max(distances)$\;
    \If{$tmp\_dist\ \textless\ distances[tmp\_max]$}{%
        $distances[tmp\_max]\gets tmp\_dist$\;
        $labels[tmp\_max]\gets l$\;
    }
}
return $mode(labels)$\;
\caption{K-Min Sorting Algorithm for KNN}
\label{custom_algo}
\end{algorithm}

\subsubsection{Bubble Sort}
The Bubble Sort produces a partially sorted array in order to take advantage of the limited number of neighbors required to produce a high accuracy using KNN, and only sort the minimum number of values defined by \emph{k}. This is done by implementing \emph{k} comparison steps or bubbles. Since the bubbles that traverse the input array only compare adjacent element, each separates a single minimum value from the rest of the input data \cite{7160066}. This process can be easily parallelized since each of the comparison bubbles can run independently of one another, and can sort elements simultaneously. The small bubble count and easy parallelism makes this implementation of Bubble Sort an effective option for FPGA platforms \cite{7160066}.

\subsubsection{Merge Sort}
There exist several Merge Sort variations that are known to be effective when run on parallel platforms \cite{9155623}. While some combine the design concepts of other sorting techniques like bitonic sort or quick sort, their function is necessarily the same and reflects the basic Merge Sort model. Resource utilization can be high for parallel Merge Sort algorithms, but their speed makes them effective for certain applications, including time-sensitive edge computing IoT systems for example DDoS detection \cite{9155623}. The Merge Sort variation implemented here follows the traditional model. The incoming data is continually bisected and each of the resulting parts is sorted and merged to produce the desired fully sorted array.

\subsubsection{Enumeration Sort}
Enumeration sort has the potential to be a good fit for the parallel structure of the FPGA fabric. This algorithm determines data points' positions in a final sorted array by comparing each element to every other. Memory accesses can be minimized since elements are not swapped or moved multiple times, making this sorting algorithm easily parallelizable \cite{1675943}.

\subsubsection{Odd-Even Sort}
Odd-even sort has the potential to reduce memory accesses and improve performance for KNN \cite{6612389}. This algorithm consists of two phases, an even phase and an odd phase, in which pairs of elements in a dataset are compared and swapped. By alternating between even and odd pairs of data each comparison step can be parallelized. However, arranging Odd-Even Sort in this manner increases its space complexity greatly, making area the primary limitation with this algorithm.

\subsubsection{The proposed K-Min Sort}     

In order to outperform the previous algorithms, we propose a custom sorting algorithm called K-Min Sort.
As seen in Algorithm \ref{custom_algo}, the distances between the test query and each element of the training data are calculated and evaluated in the same loop iterations.
Of these distances only \emph{k} are recorded, which reduces the sorting task while maintaining KNN's accuracy, and the input dataset only needs to be traversed once.
By combining the distance metric and sorting steps we minimize the run-time complexity and number of memory accesses required to retrieve and store the data and distances.
This combined step can be additionally pipelined on the FPGA as another performance benefit.
In place of a separate sorting step, Algorithm \ref{custom_algo} only compares the distance of each training data point to the furthest of the recorded neighbors at a given point in time.
As such, the number of comparison operations is minimized, proportional to the size of the input dataset. 

Despite their similarities, K-Min sort performs fewer comparisons and memory accesses than the Bubble Sort algorithm since it does not perform swap operations on each element of the input dataset.
Both algorithms traverse the training distances dataset, only sort \emph{k} distance, and can be parallelized in a similar manner, but  K-Min sort does so with a lower resource overhead.
The Merge, Enumeration, and Odd-Even algorithms each sort all of the training data point distances in order to select the \emph{k} values necessary for KNN after, but by parameterizing \emph{k} K-Min and Bubble Sort avoid this excess computation.

\subsection{HLS Simulation}
High-level synthesis (HLS) interprets and implements high level language algorithms in register transfer language (RTL) for specific hardware platforms. HLS is capable of efficiently optimizing algorithms for particular hardware systems, enabling developers to work at a high level of abstraction.
This work makes use of the HLS system in the Vivado Design Suite \cite{xilinx}.
Since we accelerate KNN using an FPGA, we need to be able to control the HLS synthesis process to make the most of the available resources on the low-cost hardware platform. To this end, Vivado provides tools for kernel optimization, function inlining, pipelining, loop unrolling and optimization, and array optimization. Finally, our results are extracted in a physical implementation as well as simulation results from HLS.

\section{Result and Discussion}\label{ref:results}
\begin{figure}[!t]
\includegraphics[width=3.5in]{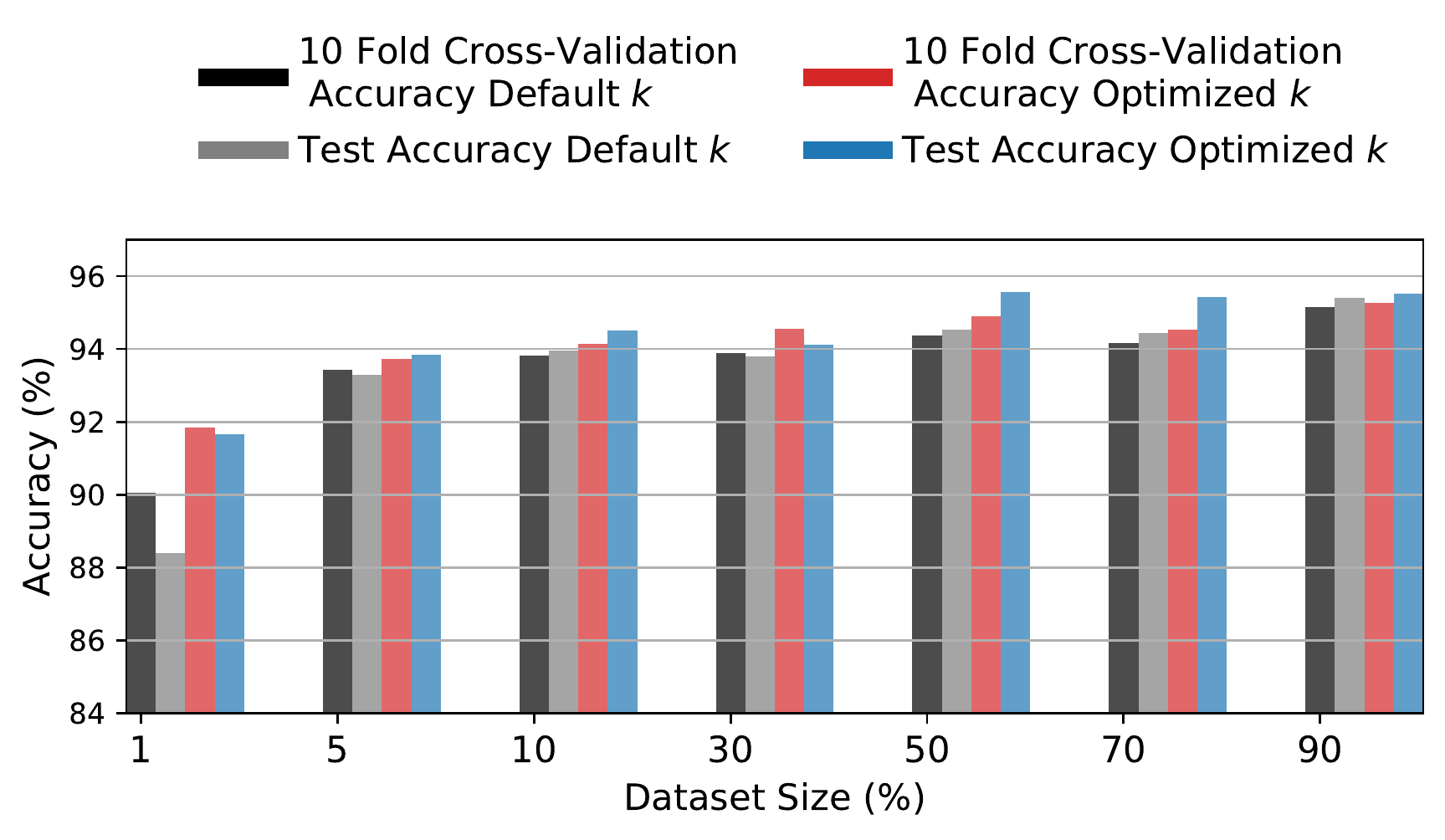}
\caption{ For different training dataset sizes, the KNN model is tuned to identify the optimal value of $k$ hyperparameter}
\label{kfold_acc}
\end{figure}

\begin{figure}[!t]
\includegraphics[width=3.5in]{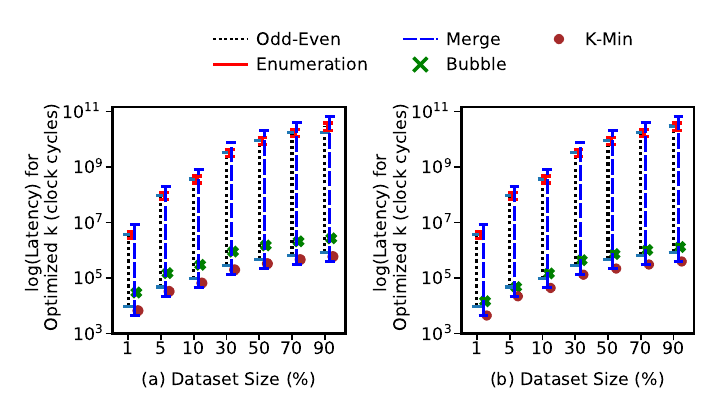}
\caption{ HLS  minimum and maximum latencies for various algorithm across different training dataset sizes of the KNN model with the default and optimal values of hyperparameter $k$}
\label{def_hls}
\end{figure}

We utilize the KNN implementation in the Scikit-learn Python library to generate test and train datasets with various size at 1\%, 5\%, 10\%, 30\%, 50\%, 70\%, 90\%  splits \cite{pedregosa2011scikit}. 
The same tool is used to run a 10-fold cross validation (CV). 
The 10-fold CV accuracy is presented in Figure \ref{kfold_acc} alongside the test accuracy for both the default $k$ value (i.e, $k$ = 5) and optimized $k$ values.
Figure \ref{kfold_acc} shows that as the training dataset size increases, the 10-fold CV and test accuracy increases.
Further, as seen in Figure \ref{kfold_acc} the accuracy exceeds 95\% at the 50\% dataset split for the default \emph{k} value, and is maintained with the reduced, optimal \emph{k}.
By optimizing \emph{k}, the amount of calculations required by KNN are minimized without a loss in accuracy.
The optimized \emph{k} value for the different dataset sizes is then implemented and tested in HLS on the FPGA platform.
Figure \ref{def_hls}(a) and \ref{def_hls}(b) show the ranges of execution times of KNN for the 5 tested sorting algorithms.
The data is gathered from HLS C synthesis reports, which approximate each algorithm's latency with high accuracy as the train-test split percentage changes.
The algorithms in Figure \ref{def_hls}(a) are compiled in HLS with the default $\emph{k}=5$, while Figure \ref{def_hls}(b) is compiled in HLS using optimal values for \emph{k}.

When an HLS C synthesis report is generated, the Vivado system evaluates the critical path of an implementation and determines its latency from the known execution times of common operations, including data manipulation, memory accesses, and movement.
However, HLS does not attempt to calculate latency values when a single critical path cannot be isolated, which may occur as a result of embedded loops and memory dependencies.
Nevertheless, the $loop\_tripcount$ pragma enables HLS to estimate the minimum and maximum latencies when a critical path cannot be immediately found by the compiler. Our work provides HLS simulation times combined with FPGA execution times in an Artix-7 FPGA. The HLS Simulation results are shown in Figure \ref{def_hls}, and the resource utilization for the algorithms and the FPGA resource consumption are shown in Tables \ref{HLS_Resource} and \ref{c_fpga_perfor}. Furthermore, our results are compared to CPU execution times for a x86 processor with two cores and four threads on a 2017 MacBook Air to compare performance in Table \ref{c_fpga_perfor}. The small embedded Artix-7 is able to achieve significant increases in performance even against a multi-core and multi-threaded CPU implementation. 

As Figure \ref{def_hls}(a) shows, the K-Min KNN solution performs very well with a minimum latency of 6.5$\times 10^3$ clock cycles at the 1\% dataset size and 580$\times 10^3$ at 90\%.
It lies just above the minimum latency of the Merge Sort algorithm at each dataset size, differing by a few percent.
The HLS synthesis performance report lists the same value for this algorithm's minimum and maximum latencies at each percentage because it is well suited to the FPGA platform and is parallelized easily in HLS. 
In Figure \ref{def_hls}(b), we see that the K-Min sorting solution exhibits a 33\% improvement in performance when the \emph{k} parameter is optimized.
This places the K-Min algorithm just below the minimum latency of Merge Sort at each dataset size.
We see a similar case in Figure \ref{def_hls}(a) when looking at Bubble Sort.
Like the K-Min solution, Bubble Sort is highly parallelizable and its performance is proportional to the size of the input dataset, so its minimum and maximum latencies are equivalent.
At a 1\% dataset size, it takes 30$\times 10^3$ clock cycles, and 3$\times 10^6$ at 90\%.
Bubble Sort lies closer to the K-Min solution in Figure \ref{def_hls}(b) and experiences a 53\% increase in performance, a greater increase than all the other algorithms.

In Figure \ref{def_hls}(b), we observe that the best performing algorithm according to minimum latencies is the K-Min solution while Bubble Sort follows in third.
The Bubble Sort algorithm performs more slowly than the K-Min solutions as a result of the comparison and swap operations it requires.
Bubble Sort stores more input data in memory than the K-Min solution, and the greater number of memory accesses that occur as a result slows down the KNN algorithm.
This limitation of memory accesses and storage is one of the primary factors that the K-Min solution sought to remedy.

When looking at the performance of Merge Sort in Figure \ref{def_hls}(a) and Figure \ref{def_hls}(b), we see that optimizing \emph{k} does not have a significant impact on this algorithm, reflecting a change of less than 0.1\%.
The minimum and maximum latencies at the 1\% dataset size are around 4$\times 10^3$ and 3$\times 10^5$ clock cycles in both figures, and are 8$\times 10^6$ and 6$\times 10^{10}$ clock cycles at the 90\% dataset size.
The HLS C synthesis report provides a range of latencies due to the complexity of the algorithm.

The latency data for Odd-Even Sort is similar to that of Merge Sort; as seen in Figure \ref{def_hls}(a) and Figure \ref{def_hls}(b), optimization does not significantly impact its performance.
At the 1\% dataset size the minimum and maximum latencies are about 9$\times 10^3$ and 3$\times 10^6$, and at the 90\% dataset size they range from 8$\times 10^5$ to 3$\times 10^{10}$ clock cycles. Like with Merge Sort the HLS C synthesis report produces ranges of latencies due to its complexity.

Merge sort's divide and merge steps are easily parallelized, but require the full input dataset to be stored in local memory.
Odd-even sort is similar, its alternating sorting steps are suitable for parallel platforms but require lots of  resources.
It is possible to minimize memory use with small datasets for both algorithms, however this is not applicable to machine learning applications and algorithms like KNN, doubly so here as a result of the limited resources of the low-cost FPGA platform \cite{9155623}\cite{6612389}.
Additionally, both do not take advantage of the \emph{k} parameter to reduce sorting operations, slowing KNN significantly.
The Vivado HLS system ensures a high level of optimization wherever possible despite the limitations of hardware platforms, but their shortcomings prevent additional performance improvements.

Among the 5 sorting algorithms, Enumeration Sort performs consistently worst.
As seen in Figure \ref{def_hls}(a) and Figure \ref{def_hls}(b), the performance of this sorting algorithm is the least affected by optimizing \emph{k}: at the 1\% dataset size the minimum and maximum latencies are 3$\times 10^6$ and 4$\times 10^6$, and at 90\% they are 2$\times 10^{10}$ and 4$\times 10^{10}$ clock cycles in both cases.
Despite the relative simplicity of the algorithm, HLS reports ranges of latencies due to its poor parallelizability on low resource FPGA platforms.

Like Merge and Odd-Even Sort, Enumeration Sort performs far better in small applications where it can be better parallelized \cite{6747409}, and fails to take advantage of KNN’s \emph{k} parameter.
It has lower resource overhead than Merge Sort, which is likely why it performs better than the worst case Merge Sort latencies. But in the best case, Merge Sort is the fastest of the three, with Odd-Even Sort following in second.
However, K-Min and Bubble Sort outperform other algorithms, and are better suited to FPGAs.
In the best case scenario where no sorting is required, all except Enumeration Sort only verify the position of each element of the input dataset.
Since the Enumeration algorithm must compare each element of the input dataset against every other element, it still has a high minimum latency.
\begin{table}[!tb] 
 \centering
\def\arraystretch{1.2}
    \caption{Latency of sorting algorithm on CPU and FPGA for optimal $k$ with 50\% training data}
        \begin{tabular}{|p{2.6cm}|p{2cm}|p{2cm}|}
        \hline
               Sorting Algorithm    & CPU (ms)      & FPGA (ms)  \\ \hline         \hline
                Odd-Even            &  1927200    & 	297615	   \\ \hline
                Enumeration         & 	1209000   & 	 281750    \\ \hline
                Merge	            &  150000     &    6024	   \\ \hline
                Bubble	            & 34811       & 	13.041	   \\ \hline
                \textbf{Proposed K-Min}               & 	\textbf{32519} 	  &   \textbf{3.913}      \\ \hline
\end{tabular}
\label{c_fpga_perfor}
\end{table}
The trivial lower bounds of both K-Min and Bubble Sort's execution times are strongly tied to the size of the input dataset while 
the additional operations of Merge, Odd-Even, and Enumeration Sort hold more weight in their performance. 
This is why the minimum latencies of Merge and Odd-Even Sort in Figure \ref{def_hls} are close to K-Min and Bubble Sort$-$the best case reflects their capacity for parallelizing more operations$-$but also have the highest latencies. 
Enumeration sort demonstrates this in the extreme by performing poorly with or without optimizing \emph{k}. 
Additionally, optimization in Figure \ref{def_hls}(b) reduces the amount of data that is processed for K-Min and Bubble Sort, enabling K-Min sort to outperform Merge Sort in the best case.
Bubble Sort and Odd-Even Sort follow Merge Sort with a 35\% difference between Bubble and Odd-Even Sort and a 54\% difference between odd-even and Merge Sort.

\begin{table}[!tb] 
 \centering
\def\arraystretch{1.2}
    \caption{HLS Resource Usage for Optimal $k$ with 50\% training data}
        \begin{tabular}{|p{2.5cm}|p{.8cm}|p{.8cm}|p{.8cm}|p{.8cm}|}
        \hline
                Sorting Algorithm   & BRAM  & DSP    & FF     & LUT  \\ \hline \hline
                Odd-Even            &  258  & 	5	 & 4084   & 3844 \\ \hline
                Enumeration         &  642  & 	5    & 3957   & 4204 \\ \hline
                Merge	            &  514	&   5	 & 4372   & 4447 \\ \hline
                Bubble	            &  258  & 	7	 & 3880   & 3628 \\ \hline
                \textbf{Proposed K-Min}               & 	 \textbf{4}	&   \textbf{5}	 & \textbf{3886}	  & \textbf{3663} \\ \hline
\end{tabular}
\label{HLS_Resource}
\end{table}

Finally, when comparing the real run times on the Artix-7 FPGA with simulation results, K-Min performs the best with an inference time of 3.913 ms as seen in Table \ref{c_fpga_perfor}.
The performance gains of the other sorting algorithms is as expected.
Due to the customization of the K-Min algorithm for an FPGA workload, our solution is able to obtain extremely optimized millisecond level performance, several orders of magnitudes faster, while only consuming $2.67\%$ of the total Block Random Access Memory (BRAM), $4.17\%$ of the DSP slices, $5.96\%$ of the total Flip Flops (FF), and $11.24\%$ of the total Look Up Table (LUT) space as seen in Table \ref{HLS_Resource}.
Therefore, this implementation is not only extremely fast, but also utilizes minimum resources on the FPGA to provide an optimal design achieving $95\%$ accuracy at 3.9 millisecond speed.





\section{Related Work}\label{ref:related_work}

FPGA platforms are cost effective and versatile for IoT and edge computing applications \cite{7301601}.
They are suitable for these tasks because they maintain steady throughput regardless of workload, have a high capacity for spatial and temporal parallelism, and are very energy efficiency \cite{article}.

In existing work we see robust FPGAs used in large scale server-side applications to supplement network security.
FPGA's speed makes them useful for real-time network intrusion detection and device identification in closed critical industrial networks \cite{8969630}. For example, a proposed k-means k-modes clustering architecture in \cite{8826329} implements highly configurable input parameters with interconnected blocks to reduce the need for reconfiguration.
\cite{8892078} allows network parameters to specify pruning, data quantization, and precision without reconfiguration at run-time.


Acceleration techniques can be refined by evaluating KNN variations to find the most suitable for certain applications, but there is a lack of existing work comparing parallel KNN variations on FPGAs.
This lack of exploration is similarly seen of other prevalent ML algorithms whose acceleration could benefit IoT applications.
The range and variety of optimization techniques made available by the parallelism of FPGAs and HLS reinvigorates KNN, whose improvements may benefit other algorithms as well.

Device identification has been accelerated with the use of fixed-point arithmetic \cite{8268745} by using 16 and 8 bit precision to reuse hardware and reduce space and resource consumption.
This also improves memory access efficiency for algorithms on FPGAs; 
it allows distant neighbors to be discarded early in KNN \cite{8988717}.
This modification is particularly useful for handling large datasets on FPGAs with few memory resources.

While these existing existing parallel acceleration techniques are informative, they are not designed for fast, real-time edge computing applications, 
but rather for server applications and the workloads of large scale data centers \cite{9034938}. 
This absence of research regarding KNN for edge computing and IoT applications extends to the parallelization of sorting algorithms \cite{7160066} \cite{8623861}.
Only niche applications of KNN acceleration exist for TCP/IP packet inspection \cite{8674974}, NIC outlier filtering \cite{7912620}, and flow-based network traffic attack detection \cite{article2}.

This work expands on existing research, and accelerates the computationally intensive sorting and selecting task of KNN with a low-cost FPGA suitable for edge computing platforms.
Traditional sorting algorithms like Merge \cite{9155623} and Bubble Sort \cite{7160066} are not feasible on small FPGAs.
Therefore, to get the best performance and efficiency out of our low-cost system, we make use of several acceleration techniques with a custom sorting algorithm to develop a KNN solution that is well-suited to small-scale IoT networks.

\section{Conclusion}\label{ref:conclusion}
The proliferation of IoT devices requires increased flexibility for home network management and more lightweight and robust security solutions.
Software-Defined Networking (SDN) and Machine Learning (ML) have been proposed as efficient solutions to provide device classification and malicious traffic detection.
However, Field Programmable Gate Arrays (FPGAs) can further enhance this efficiency due to their fine-grained parallelism and parallel processing efficiency.
This is demonstrated by the proposed FPGA-based KNN algorithm classifying a network flow with $95\%$ accuracy in 4 milliseconds compared to the required 57 seconds on a CPU-based platform.
This stark performance difference is indicative of the potential of FPGAs to benefit other ML algorithms which would otherwise be dismissed on CPUs due to the constraints of serial processing platforms.
Our work hopes to encourage further research in using FPGA technology in edge networking applications due to its speed and energy efficiency.

\bibliography{IEEEabrv.bib}
\bibliographystyle{IEEEtran}

\end{document}